# Electrical generation and detection of terahertz signal based on spin-wave emission from ferrimagnets

Zhifeng Zhu[1,2,*], Kaiming Cai[1], Jiefang Deng[1], Venkata Pavan Kumar Miriyala[1], Hyunsoo Yang[1], Xuanyao Fong[1], Gengchiau Liang[1,†]

[1]Department of Electrical and Computer Engineering, National University of Singapore, Singapore 117576

[2]School of Information Science and Technology, ShanghaiTech University, Shanghai, China 201210

Terahertz (THz) signals, mainly generated by photonic or electronic approaches, are being sought for various applications, whereas the development of magnetic source might be a necessary step to harness the magnetic nature of electromagnetic radiation. We show that the relativistic effect on the current-driven domain-wall motion induces THz spin-wave emission in ferrimagnets. The required current density increases dramatically in materials with strong exchange interaction and rapidly exceeds $10^{12}$ A m$^{-2}$, leading to the device breakdown and thus the lack of experimental evidence. By translating the collective magnetization oscillations into voltage signals, we propose a three-terminal device for the electrical detection of THz spin wave. Through material engineering, wide frequency range from 264 GHz to 1.1 THz and uniform continuous signals with improved output power can be obtained. As a reverse effect, the spin wave generated in this system is able to move ferrimagnetic domain wall. Our work provides guidelines for the experimental verification of THz spin wave, and could stimulate the design of THz spintronic oscillators for wideband applications as well as the all-magnon spintronic devices.

# I. Introduction

The generation of terahertz signals (THz), approximately from 300 GHz to 3 THz, is of great interest due to the extensive applications in biology, security, material characterization, and high speed wireless communication [1-4]. Since the THz spectrum lies between two well-established domains (that is microwaves and light), it can be generated indirectly using solid-state electronic devices combined with the frequency multiplier [5-7] or the photonic approaches using the quantum cascade laser [8,9], requiring additional peripheral circuits for the frequency conversions. On the other hand, the direct generation using Josephson junctions [10,11] and resonant-tunneling diode [12-14] are constrained by the cryogenic cooling and large device size (i.e., due to the capacitor structure), respectively.

In contrast to the electronic and photonic means, the strong exchange interaction (> 10 tesla [15]) in antiferromagnets (AFMs) offers room-temperature intrinsic THz frequency [16-19]. It has been theoretically predicted that the domain-wall (DW) velocity in AFMs is limited by the maximum spin-wave group velocity ($v_{g,max}$) [19]. As the DW velocity ($v_{DW}$) approaches $v_{g,max}$, its increment becomes smaller, and the excessive energy provided by the driving force is dissipated in the form of THz spin-wave emission. Although the abundant AFMs are good platforms for theoretical studies, the complete magnetization cancellation prevents experimental determination of magnetic states. By contrast, the ferrimagnets (FiMs) [20-23] with antiparallel exchange coupled rare earth (RE) and transition metal (TM) alloys possess finite magnetizations due to the unequal sublattices, and many interesting phenomenons have been predicted in FiMs, such as the self-focusing skyrmion racetracks [24], and the fast DW motion without Walker breakdown [21,25-27]. Therefore, the FiMs are suitable materials for developing ultrafast spintronic devices by taking advantages of both ferromagnets (FMs) and AFMs properties.

In this article, we theoretically study the current-driven DW motion in the FiM[Gd$_x$(FeCo)$_{100-x}$]/heavy metal tungsten (W) bilayer (see Fig. 1(a)), where $v_{DW}$ firstly increases and then saturates at large current density ($J_c$), in contrast to the linear trend predicted by the analytical model based on the collective coordinate approach [28], which describes the DW motion in one dimension using two variables (i.e., the location and magnetization angle of the DW center) and assumes a rigid DW profile. We then show that this deviation can be corrected by applying the Lorentz contraction on the DW width ($\lambda$) to incorporate the upper bound of $v_{g,max}$ due to the relativistic effect. In addition, the DW motion in the velocity saturation region is accompanied by the emission of THz spin waves. Although many experimental studies on the current-driven DW motion in FiMs have been conducted [25-27,29,30], evidence of spin-wave emissions has not been identified. To understand the conditions of spin-wave emissions, the parametric effects are studied, and we find that the critical current density ($J_{sw}$) required to excite the spin wave increases dramatically with the exchange constant ($A_{ex}$) and exceeds $10^{12}$ A m$^{-2}$, which can lead to device breakdown and thus prevents experimental verifications. To excite THz spin waves at small $J_{sw}$, the suitable material requires small $A_{ex}$, large crystal anisotropy ($K$), and large Dzyaloshinskii-Moriya interactions constant ($D$). Moreover, we propose a three-terminal device structure to electrically detect the THz signal, where the collective magnetization oscillation is translated into voltage signals through the tunnel-magnetoresistance (TMR) effect. Finally, we show that a uniform continuous signal with improved output power can be obtained by reducing the damping constant ($\alpha$), and a high speed DW motion with $v_{DW} = 1.5$ km s$^{-1}$ is predicted. These results could provide insights for the experimental investigation of THz spin wave, fast racetrack memory, and applications using the THz spintronic oscillator.

## II. Methods

The spin-orbit torque driven domain-wall motion in FiM is modelled using the one-dimensional atomistic model [31-33], which includes antiferromagnetic coupled TM and RE elements. The Hamiltonian is given by $E_i = A_{ex}\sum_i \mathbf{S}_i \cdot \mathbf{S}_{i+1} - K_j \sum_i (\mathbf{S}_i \cdot \hat{\mathbf{z}})^2 + \kappa_j \sum_i (\mathbf{S}_i \cdot \hat{\mathbf{x}})^2 + D\sum_i \hat{\mathbf{y}} \cdot (\mathbf{S}_i \times \mathbf{S}_{i+1})$, where $\kappa_j$ is the domain-wall hard-axis anisotropy. The spin dynamics of each sublattice is described by the atomistic Landau-Lifshitz-Gilbert equation $\partial \mathbf{S}_i/\partial t = -\gamma_i \mathbf{S}_i \times \mathbf{B}_{\text{eff},i} + \alpha_i \mathbf{S}_i \times \partial \mathbf{S}_i/\partial t - \gamma_i \hbar J_c \theta_{\text{SH}}/(2eM_{\text{S},i}t_{\text{FiM}})[\mathbf{S}_i \times (\mathbf{S}_i \times \hat{\mathbf{y}}) - \beta \mathbf{S}_i \times \hat{\mathbf{y}}]$, where $\hbar$ is the reduced Planck constant, $\theta_{\text{SH}}$ is the spin-Hall angle, $e$ is the electron charge, $M_{s,i}$ is the saturation magnetization, $t_{\text{FiM}}$ is the thickness of the ferrimagnetic layer, $\gamma_i = g_i\mu_B/\hbar$ is the gyromagnetic ratio where $g_i$ is the $g$-factor and $\mu_B$ is the Bohr magneton, $\mathbf{B}_{\text{eff},i} = -(1/\mu_i)\partial E_i/\partial \mathbf{S}_i$ is the effective field where $\mu_i = M_{s,i}d^3$, and $\beta$ denotes the ratio between the field-like torque (FLT) and damping-like torque (DLT). The four terms on the right-hand side are precession, damping, DLT, and FLT, respectively. The parameters used in our simulation are summarized as follows. $A_{ex}$ = 3 meV, $K_{\text{TM}}$ = $K_{\text{RE}}$ = 0.04 meV, $\kappa_{\text{TM}}$ = $\kappa_{\text{RE}}$ = 0.2 μeV, $D$ = 0.128 meV, $\alpha_{\text{TM}}$ = $\alpha_{\text{RE}}$ = 0.015, $t_{\text{FiM}}$ = 0.4 nm, $g_{\text{TM}}$ = 2.2, $g_{\text{RE}}$ = 2, $\theta_{\text{SH}}$ = 0.2, $\rho_{\text{Ta}}$ = 200×10$^{-8}$ Ω m, $\rho_{\text{FiM}}$ = 248×10$^{-8}$ Ω m [34], $TMR$ = 100% [35], $R_{\text{P}}$ = 500 Ω [36,37]. The thickness of the W and FiM layer are 5 nm and 0.4 nm, respectively.

The temperature, which can be tuned to change the material from Gd to FeCo dominated regions, is an important parameter in FiMs. The effect of temperature is generally taken into account by describing the change of magnetization using a power-law [27,38,39]. Recent studies [27,40] show that the temperature change induced by Joule heating might change the net magnetization from the Gd to FeCo dominant. However, in this study, we assume a negligible effect from Joule heating by limiting $J_c$ below $10^{12}$ A m$^{-2}$ and choosing the substrate with high cooling efficiency [41,42]. Therefore, we focus on the magnetization dynamics in the FiM at a stable temperature of 300 K with $M_{s,TM} = 1149\times10^3$ A m$^{-1}$ and $M_{s,RE}$ =$1012\times10^3$ A m$^{-1}$.

**III. Device structure and THz spin-wave generation**

As shown in Fig. 1(a), a bilayer structure with the FiM deposited on top of the tungsten layer is first studied to understand the spin-wave generation. With sufficient Dzyaloshinskii-Moriya interactions (DMI), the initial magnetic state in the FiM is Néel wall with right-handed chirality due to the positive sign of DMI in W (Note 1 in [43]), in agreement with ref. [44]. $J_c$ in the $x$ direction generates vertical spin current polarized along the $y$ axis, and the resulting spin-orbit torque (SOT) efficiently moves the DW in one direction depending on the current polarity. According to the theoretical study based on the collective coordinate approach [28], $v_{DW}$ of a stable Néel wall is a linear function of $J_c$ as

$$v_{DW}=(s_{FeCo}+s_{Gd})\pi\lambda B_D/(4\alpha), \qquad (1)$$

where $s_{FeCo(Gd)} = M_{s,FeCo(Gd)}/\gamma_{FeCo(Gd)}$ is the angular momentum for FeCo(Gd), and $B_D = \hbar\theta_{SH}J_c/(2et_{FiM}s_{FeCo}s_{Gd})$ is the effective SOT field. In contrast to the straight line predicted by equation (1), numerical simulations using the atomistic model (see Fig. 1(b)) show an increase and saturation trend. For $J_c > J_{sw}$, the spin wave emerges at the DW and vanishes in the direction opposite to the DW motion (see Note 2 in [43]). The frequency of the spin wave is identified in the THz range by performing fast Fourier transform (FFT) on the magnetization. The spatial profile of the spin wave is shown in Fig. 1(c), where the DW is located at 148 nm and moves against the electron flow [44]. On the left side of the DW, $m_x$ and $m_y$ show strong spatial variations, corresponding to the atoms precessing at different phase and amplitude as schematically depicted in Fig. 1(a). The spatial profile of spin waves is related to the generation method. For example, the spin wave with uniform amplitudes can be generated by applying microwave field to the whole sample [45], whereas the one generated by spin-torque nano-oscillators [46-48] has the largest amplitude at the current injection point and propagates to the surroundings with reduced amplitude due to the inevitable damping. Similar to the second case, the source of spin waves in this study is located at the DW, and the oscillation amplitude reduces in atoms which are far away from the source. Therefore, the generation of spin waves reduces the DW energy, inhibiting the linear increase of $v_{DW}$ as a function of $J_c$. Theoretical studies have identified that $v_{DW}$ is limited by $v_{g,max}$ due to the relativistic effect [19,21], which is supported by the result that the velocity trend shown in Fig. 1(b) can be well explained by adding the Lorentz contraction into the collective coordinate approach (see more discussions in Fig. 3). Furthermore, we have numerically verified that the spin wave generated in our device is able to move another chiral DW in FiM. Similar phenomenon has been predicted in FM [49] and AFM [50]. The interplay between spin wave and DW [51-55] can then be explored to optimize the racetrack memory.

Despite these theoretical predictions and numerical results, no direct evidence of spin-wave emission has yet been experimentally observed [25-27]. Therefore, we investigate the conditions for spin-wave emission by studying the dependence of $J_{sw}$ on material parameters. As shown in Fig. 1(d), $J_{sw}$ increases dramatically and rapidly surpasses $10^{12}$ A m$^{-2}$ for $A_{ex} > 8$ meV, and this trend remains the same when $K$ or $D$ is changed (shown in Note 3 in [43]). The large $J_{sw}$, as a primary obstacle in experiments, can easily lead to sample breakdown. In addition, since the spin current generated by the spin-orbit coupling is polarized in the $y$ direction, large $J_c$ would distort the Néel wall, resulting in inefficient DW motion [19]. Therefore, the sample with small $A_{ex}$, large $K$, and large $D$ is required for a stable Néel wall with small $J_{sw}$, which can facilitate the experimental exploration of spin-wave emission.

**IV. Electrical detection using a three-terminal structure**

Spin waves are mainly detected using microstrip antennas [56] or optical approaches [57], which are unsuitable for the integrated devices. In contrast, the giant magnetoresistance (GMR) and TMR are widely used in spintronics for detecting magnetization states by measuring voltage signals [58-64]. Limited by the lateral size of magnetic stacks (> 30 nm), the collective spins with nonuniform amplitude have to be utilized (as shown in Fig. 1(c)). Using the three-terminal structure schematically illustrated in Fig. 2(a), we show that the THz spin wave can be electrically detected, and the frequency of the output signal is identical to that of the single spin. The generation of spin wave has been discussed in the previous section, and the detection of magnetization states can be realized by passing a small current ($I_{MTJ}$) through the magnetic tunnel junction (MTJ) consisting of the FM, MgO, and FiM layers. When the DW passes through the MTJ, the dynamics of collective spins are manifested as the change in resistance due to the TMR effect, resulting in alternating electrical signals ($V_o$) across the MTJ. To extract the alternating component of $V_{MTJ}$, a bias-tee is used to measure the THz signal at the radio frequency (RF) port. Two independent current sources are used in two separate channels, enabling the independent control of $v_{DW}$ and $V_o$. The effect of $I_{MTJ}$ on DW motion is negligible, and $I_c$ does not directly affect $V_o$ because the vertical current is vanished (Note 5 in [43]). As a result, the output power can be enhanced by simply increasing $I_{MTJ}$. In order to correctly capture the current distributions, a distributed resistance model, as shown in Fig. 2(b), is developed and solved analytically (Note 4 in [43]). When the current distribution is obtained, the averaged current passing through the W layer is used to calculate the SOT, followed by the simulation of magnetization dynamics using the atomistic spin model (Methods). The MTJ resistance, as a function of the averaged FiM magnetization, is recomputed at each time step, and then the current distribution is calculated again. This process is repeated to get the time evolution of magnetizations. In addition, $V_o$ can also be improved using the MTJ with large TMR, which has been reported in several studies, such as 15% in the TbCoFe/CoFeB/MgO/CoFeB/TbCoFe

MTJ [36], and 55% in the GdFeCo/CoFe/Al$_2$O$_3$/CoFe/TbFeCo MTJ [35]. It is worth noting that the detailed MTJ stack is less important as long as it can provide sufficient TMR. Therefore, these FiM-based MTJs with large TMR can be used readily in this proposal to enhance the output signal.

The proof-of-concept oscillation for the averaged $m_y$ and $\Delta V_o$, defined as $V_o - 252.465$ mV, are shown in Figs. 2(c) and 2(d), respectively. A clear oscillation pattern appears when the DW passes through the MTJ at 95 ps (referred to Note 6 in [43]), and then it gradually vanishes as the DW moves further away. As discussed in Fig. 1(b), the spin wave is originated from the energy dissipation of DW. With the DW moving away, the atoms lose energy and slowly return to the stable state due to the damping. The resulting $V_o$ oscillates with a peak-to-peak amplitude of 14.5 μV. As shown in the inset of Fig. 2(d), the main frequency ($f$) of $V_o$ is 437 GHz, which is identical to that of single spin precession (cf. Note 6 in [43]), thus supporting that $V_o$ has the same origin with the spin wave. Since these atoms also experience the SOT, they precess with negligible amplitude at $f_{SOT}$. This is similar to the current-induced jiggling in FMs, which is determined by the combined field of SOT, exchange, and anisotropy (cf. the inset of Fig. 2(d) and Note 6 in [43]).

**V. Appearance of relativistic effect and performance improvement**

To elucidate the origin of the spin wave emission, we first study the relation between $v_{DW}$ and $J_c$. According to equation (1), $v_{DW}$, plotted in Fig. 3(a) using the solid line, is a linear function of $J_c$, contradicting to the saturation trend as shown in Fig. 1(b). In addition, $v_{DW}$ without Lorentz contraction at $A_{ex} = 8$ meV and $J_c = 10^{12}$ A m$^{-2}$ is 5.9 km s$^{-1}$, which is four times larger than that from the atomistic simulation. To resolve these discrepancies, it has been pointed out that the DW motion in AFMs or FiMs with large DMI requires additional consideration of the relativistic effect [19,21], which imposes the Lorentz contraction on $\lambda$, resulting in the deviation from its equilibrium value ($\lambda_{eq}$) by a factor of $\sqrt{1-(v_{DW}/v_{g,max})^2}$, where $v_{g,max}$ is obtained from the dispersion relation of the spin wave. Such correction is similar to the length measurement of a fast moving object, which, according to the special relativity, should be modified by the Lorentz factor $\sqrt{1-(v/c)^2}$ with $v$ and $c$ denoting the speed of object and the speed of light, respectively. According, the Lorentz contraction is applied to $v_{DW}$ of the current-driven DW motion in FiM, and then equation (1) is modified as

$$v_{DW} = b(s_{FeCo}+s_{Gd})\pi\lambda_{eq}\sqrt{1-(v_{DW}/v_{g,max})^2}B_D/(4\alpha), \tag{2}$$

where $v_{g,max}=8A_{ex}/[d^2(s_{FeCo}+s_{Gd})]$ with the lattice constant $d = 0.4$ nm, and $b = 0.27$ is a fitting coefficient. Assuming that $\lambda_{eq} = d\sqrt{2A_{ex}/K}$, the corrected $v_{DW}$, plotted as the dash line in Fig. 3(a), shows a clear saturation trend with the maximum $v_{DW}$ below 1 km s$^{-1}$.

In addition, the atomistic simulation predicts that $v_{DW}$ increases linearly with $A_{ex}$ (shown as the circles in Fig. 3(b)), whereas it is a square root relation according to equation (1), as shown in Fig. 3(b) using the solid line. By correcting $v_{DW}$ using the Lorentz contraction, $v_{DW}$ predicted by equation (2) (see the dash line of Fig. 3(b)) shows a good agreement with the numerical result. Due to the large $J_c = 10^{12}$ A m$^{-2}$ used in Fig. 3(b), the Néel wall is distorted and a sizeable Bloch component (i.e., $m_y$) appears in all the studied samples. It has been discussed in ref. [19] that the damping-like SOT cannot move the Bloch wall, and hence it is reasonable to use $b < 1$ to get quantitative agreements. Therefore, equation (1) from the collective coordinate approach is insufficient to explain the numerical results, and the introduction of the relativistic correction leads to excellent agreements.

Next, we study $v_{DW}$ as a function of $J_c$ at different $A_{ex}$. As shown in Fig. 3(c), high speed DW motions with a maximum $v_{DW} = 1.5$ km s$^{-1}$ appears at $A_{ex} = 8$ meV, which can be useful in applications such as the racetrack memory [65]. Similar to Fig. 1(d), where the MTJ structure and distributed resistance model are excluded, $J_{sw}$ increases dramatically with $A_{ex}$. Although there are debates on the relative amplitude of the DLT and FLT in the magnetic layer attached to the heavy metal [66-69], it is well accepted that both torques are sizeable. Using the FLT to DLT ratio $β = 1.1$, which is in accordance with the Rashba-Edelstein effect [67], we find that the FLT, albeit with larger magnitude, has negligible effect on $v_{DW}$ (cf. the third panel of Fig. 3(c)). The conclusion holds under different FLT strength or even the FLT with an opposite sign [70]. This is consistent with the study in AFM that the FLT does not affect the dynamics of Néel wall [19]. Despite the substantial Bloch component at large $J_c$, the FLT in this study is insufficient to change $v_{DW}$ [19].

In addition to the fast DW motion, the frequency of $V_o$ can be tuned in a wide range, which is advantageous over existing devices. For example, the wireless communication system at 200 to 300 GHz relies on the SiGe or Si-CMOS technology [4,71], whereas GaN, InP, or photonics devices oscillated at higher frequency are required for applications above 500 GHz [4]. The necessity of integrating different technologies complicates the transceiver design. As shown in Fig. 4(a), $f$, as a function of $A_{ex}$, can be changed in a wide range from 264 GHz to 1.1 THz, which can be intuitively understood as consequences of the increased effective field. The reduced frequency below 300 GHz may also enable experimental verification using real-time oscilloscope. Therefore, the large frequency window in the spin-wave based spintronic oscillator offers a unified platform for the THz applications.

Besides the frequency tunability, a uniform signal with large output power is also preferred to simplify the peripheral circuits [71]. Since the vanishing of spin wave is induced by the loss of energy input when the DW moves away, the uniformity of oscillation signal can be improved by either reducing $v_{DW}$ or $α$. Since $α$ is directly associated with material properties, we investigate the effect of $α$ in this study. As shown in Figs. 4(b) and 4(c), $V_o$, in which the main frequency remains the same, becomes more uniform and its magnitude is four times larger when $α$ is changed from 0.015 to 0.01. Moreover, $f_{SOT}$ becomes more manifested at smaller $α$. To understand this, the atoms precessing at different frequencies are investigated. The reduction of $α$ increases the relaxation time of all atoms, most of which are precessing at $f_{SOT}$ without emitting spin waves. As a result, the relative amplitude change is larger in $f_{SOT}$ compared to that in $f$. However, $V_o$ is still dominated by $f$ because of its larger averaged oscillation amplitude. Therefore, the engineering of material parameters can significantly improve the device performance, opening up possibilities of using spintronic oscillators in wideband applications.

**VI. Discussion and Conclusion**

In conclusion, we have studied the conditions of THz spin-wave emission in the FiM/W bilayer under the SOT. The DW velocity is limited by the maximum group velocity due to the relativistic effect, and the excessive energy is dissipated in the form of THz spin waves. The critical current required for the spin-wave emission increases dramatically in materials with strong exchange coupling, and this trend remains the same when $K$ or $D$ is changed. Therefore, the FiM with small $A_{ex}$, large $K$, and large $D$ is suitable for the experimental demonstration. We then propose the electrical detection of the THz spin wave in a three-terminal structure by translating the collective magnetization oscillation into voltage signals. We show that a wide frequency range from 264 GHz to 1.1 THz, a uniform continuous signal with improved output power, and a fast DW motion at 1.5 km s$^{-1}$ can be obtained by optimizing material parameters. This work promotes understanding of magnetization dynamics in FiM, provides candidates for the fast racetrack memory, and should stimulate the design of THz oscillator for wideband applications using a unified platform.

**Corresponding Authors:** *a0132576@u.nus.edu, †elelg@nus.edu.sg

**Acknowledgements**

This work at the National University of Singapore was supported by CRP award no. NRF-CRP12-2013-01, and MOE-2017-T2-2-114.

**Figures**

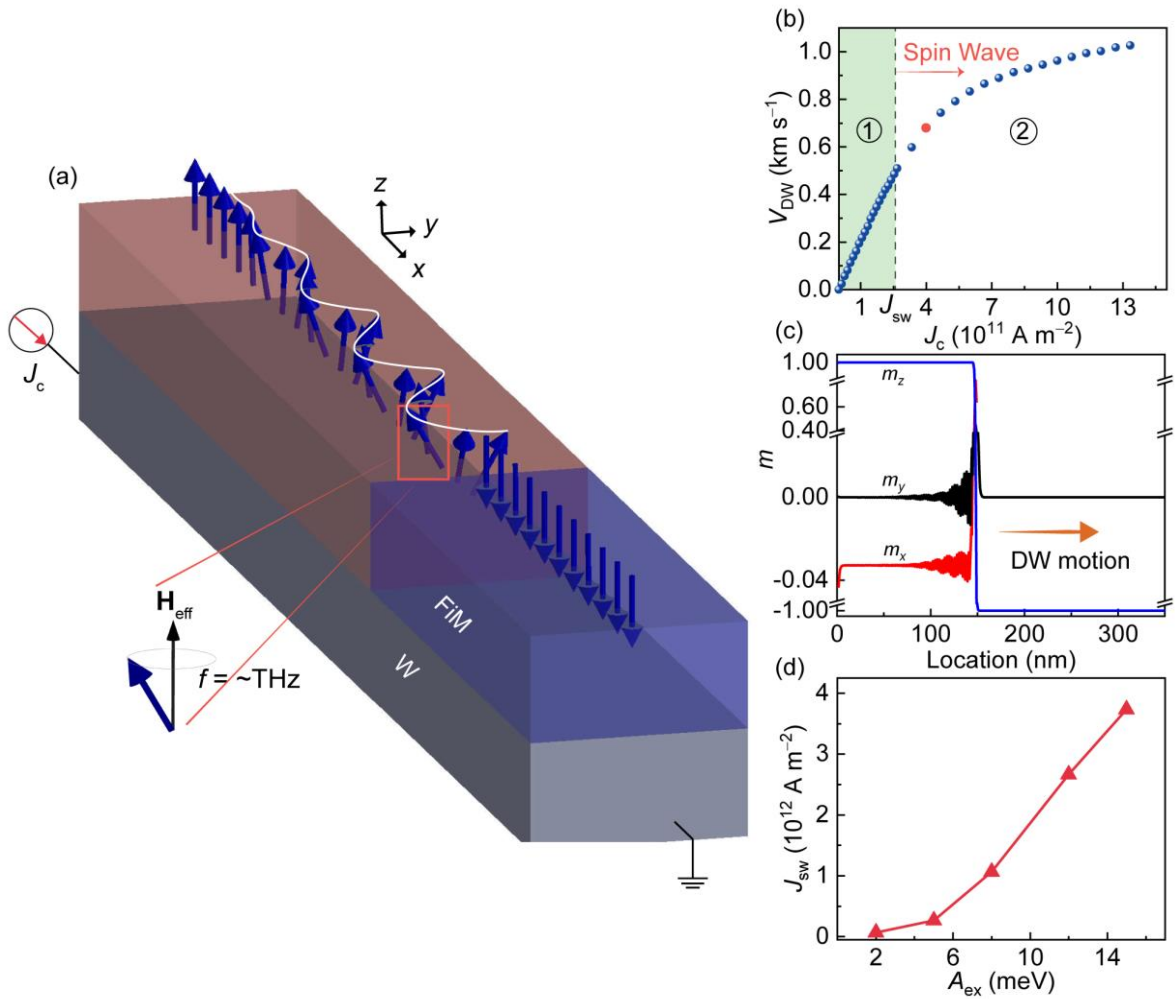

**Figure 1**. Illustration and condition for the spin wave emission. (a), Schematic view of the THz spin wave generated in the FiM/W bilayer, where the red and blue color in the FiM layer distinguish the up and down domain respectively. The length and width of W and FiM are 400 nm and 40 nm, respectively. The current moves the Néel wall along the $+x$ direction. The spin wave is initiated at the DW and vanishes in $-x$ direction. The precession frequency of each spin is identified in THz range. (b), The DW velocity as a function of $J_c$. The dash line separates regions without (region 1) and with (region 2) spin wave. (c), The spatial profile of magnetization at $J_c = 4\times10^{11}$ A m$^{-2}$ and $A = 5$ meV, where the DW is located at 148 nm. The DW moves to the right as marked by the brown arrow. (d), Critical current density for the spin-wave emission as a function of exchange constant.

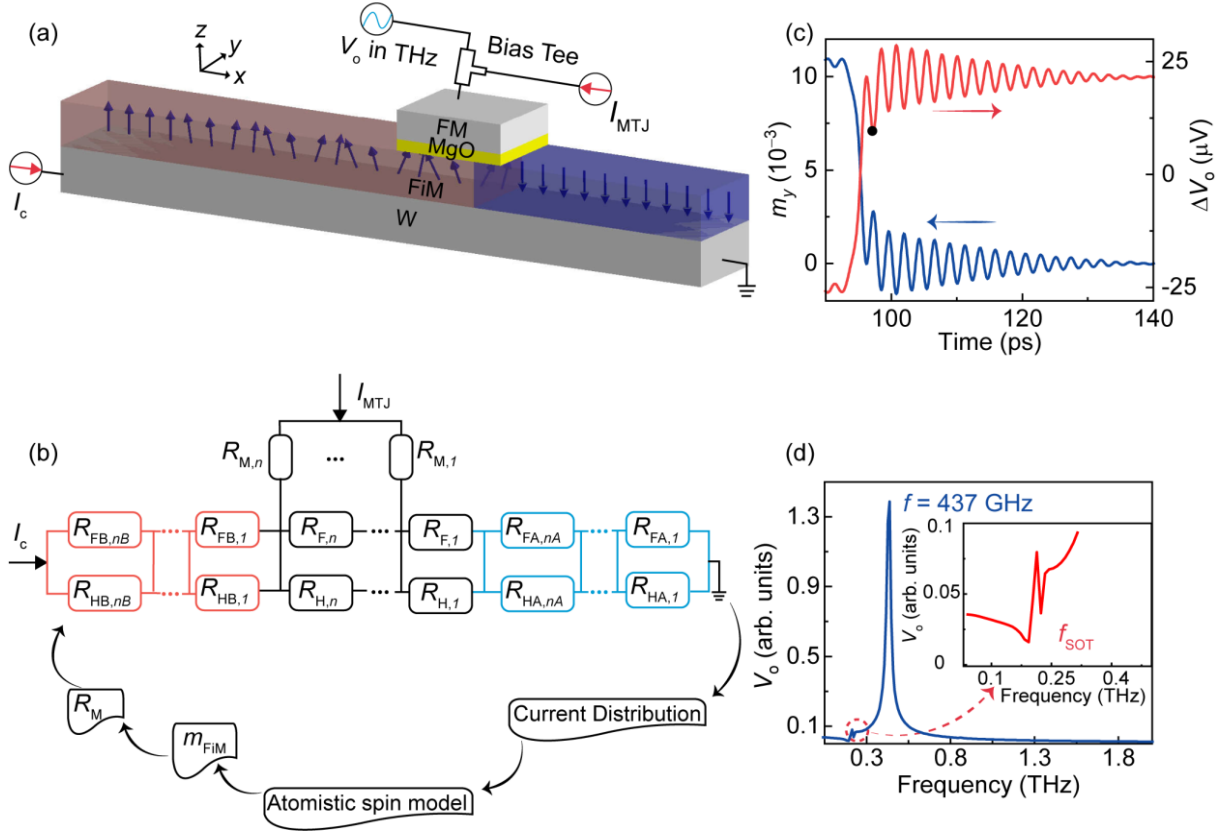

**Figure 2**. Electrical detection of the THz signal. (a), Schematics of the three-terminal structure. The length and width of both MgO and FM layer are 40 nm. The FM layer has in-plane easy axis along in the $+y$ direction. The bias tee is a three-port diplexer consisting of a capacitor and an inductor. $I_c = 160$ μA and $I_{MTJ} = 16$ μA are used in this study unless otherwise stated. (b), Simulation workflow. The current distribution is calculated using the distributed resistance model, and then the average current passing through the W layer is input to the atomistic spin model to get the magnetization dynamics. At each time step, the MTJ resistance ($R_M$) is updated based on the magnetization of the FiM layer ($m_{FiM}$), followed by the recalculation of current distribution. This process forms a closed loop and free running to get the time evolution of magnetizations. The number of segmentation $nA = 10$, $nB = 10$, $n = 13$ which gives 0.03% difference in $I_{HA}$ compared to $n = 193$ (cf. Note 5 in [43]). Refer to Note 4 in [43] for the notation of parameters. (c), Time evolution of $m_y$ and $\Delta V_o$, defined as $V_o - 252.465$ mV, at $I_c = 160$ μA and $I_{MTJ} = 16$ μA. $m_y$ is obtained by averaging spins under the MTJ. The reduction of oscillation amplitude is induced by the loss of energy input when the DW moves away. (d), The frequency spectrum for the time region starting from 97.26 ps, marked by the black dot, to 200 ps. The main frequency is originated from spin wave at 437 GHz and a smaller SOT induced precession with negligible amplitude appears at 213 GHz, showing as the inset.

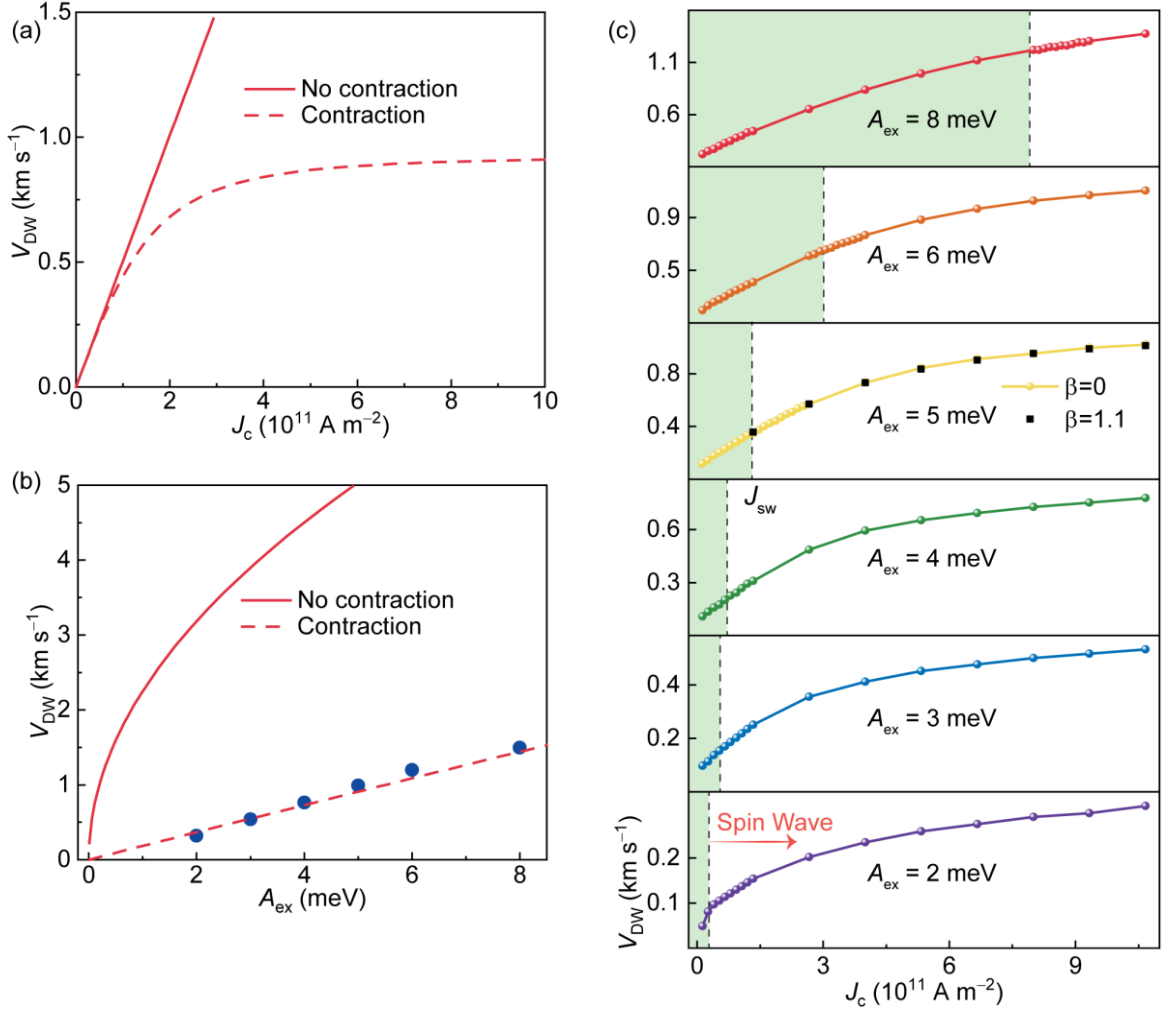

**Figure 3**. Relativistic effect on the domain-wall motion. (a), $v_{DW}$ as a function of $J_c$ at $A_{ex}$ = 5 meV for systems with (dash line) and without (solid line) the Lorentz contraction. (b), $v_{DW}$ as a function of $A_{ex}$ with $I_c$ = 160 μA and $I_{MTJ}$ = 16 μA. Results from the atomistic model are showing as the blue dots. The analytical results based on equations (1) and (2) are plotted as the solid and dash line, respectively. $A_{ex}$ above 8 meV is not considered since the $J_{sw}$ is above $10^{12}$ A m$^{-2}$, which can easily leads to device breakdown. (c), $v_{DW}$ as a function of $J_c$ at different $A_{ex}$. The dash lines define the critical current density, above which the spin waves are emitted. The effect of FLT is studied in the sample with $A_{ex}$ = 5 meV, where the black squares denote $v_{DW}$ with $\beta$ = 1.1.

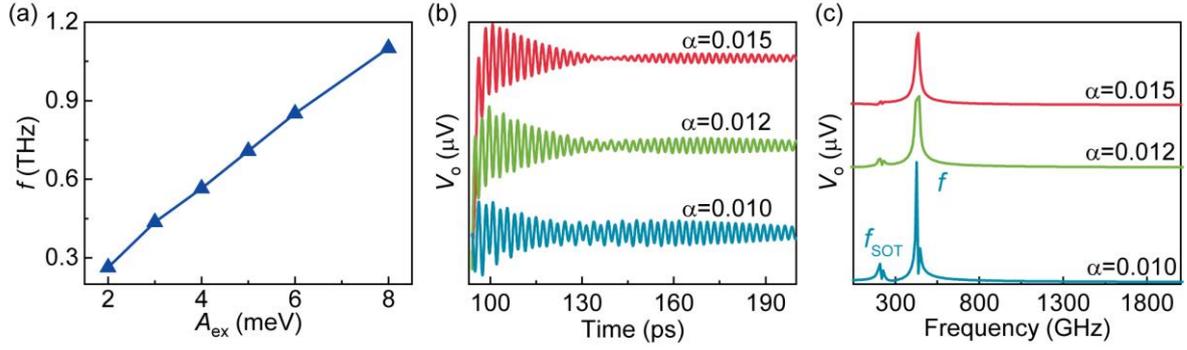

**Figure 4**. Parametric effect on the oscillator performance. (a), Numerical results of $f$ as a function of $A_{ex}$ with $I_c$ = 160 µA and $I_{MTJ}$ = 16 µA. The maximum frequency is 1.1 THz at $A_{ex}$ = 8 meV. (b, c), Time evolution (b) and frequency spectrum (c) of $V_o$ in samples with different damping constant at $I_c$ = 160 µA and $I_{MTJ}$ = 16 µA. The curves are vertically offset for clarity, i.e., 36 µV in (b) and 3.4 µV in (c). $V_o$ becomes more uniform and maintains for a longer time when the damping constant is reduced. $f$ remains nearly unchanged for different $\alpha$, i.e., $f$ = 437 GHz, 442 GHz, and 428 GHz for $\alpha$ = 0.015, 0.012, and 0.01, respectively. A substantial $f_{SOT}$ appears at 209 GHz for $\alpha$ = 0.01.